\documentclass[apj,twocolumn]{emulateapj}

\usepackage{graphics}
\usepackage{color}
\usepackage{natbib}

\usepackage{fancybox}
{\begin{Sbox}\begin{minipage}}%
{\end{minipage}\end{Sbox}\fbox{\TheSbox}}

\slugcomment{v3.3 \today~}

\newcommand{\spitzer}{{\em Spitzer}\ }
\newcommand{\spitzerirs}{{\em Spitzer}/IRS\ }
\newcommand{\spitzerirssdss}{{\em Spitzer}/IRS-SDSS\ }
\newcommand{\saga}{{\em S$^3$AGA}}

\newcommand{\sevenrm}{\rm\scriptsize}

\newcommand{\hbeta}{H{$\beta$}}
\newcommand{\halpha}{H{$\alpha$}}

\newcommand{\NII}{[N{\sevenrm\,II}]}

\newcommand       \mum        {\,{\rm \mu m}}
\newcommand       \SilAbs       {\Delta\tau_{9.7}}
\newcommand       \simali       {\,{\sim}}

\newcommand       \simlt        {\lesssim}
\newcommand       \simgt        {\gtrsim}

\newcommand       \NH       {N_{\rm H}}
\newcommand       \km        {\,{\rm km}}
\newcommand       \s        {\,{\rm s}}

\begin{document}

\title[AGN optical-to-silicate extinction]{ 
Dust in Active Galactic Nuclei: Anomalous Silicate to Optical Extinction Ratios? 
}
\author{Jianwei Lyu\altaffilmark{1,2}, 
             Lei Hao\altaffilmark{1,\dag}, 
             and Aigen Li\altaffilmark{2}
}
\altaffiltext{1}{Key Laboratory for Research in Galaxies and Cosmology,
                 Shanghai Astronomical Observatory, 
                       Chinese Academy of Sciences, 
                       80 Nandan Road, Shanghai 200030, China
                       }
\altaffiltext{2}{Department of Physics and Astronomy, 
                       University of Missouri, 
                       Columbia, MO 65211, USA
                       }
\altaffiltext{\dag}{haol@shao.ac.cn}

\begin{abstract}                                                
Dust plays a central role in the unification theory of active galactic nuclei
(AGNs). However, little is known about the nature (e.g., size, composition) of
the dust that forms a torus around the AGN.  In this Letter we report a
systematic exploration of the optical extinction ($A_V$) and the silicate
absorption optical depth ($\SilAbs$) of 110 type 2 AGNs.  We derive $A_V$  from
the Balmer decrement based on Sloan Digital Sky Survey data, and $\SilAbs$ from the {\it
Spitzer}/InfraRed Spectrograph data.   We find that with a mean ratio of $\langle
A_V/\SilAbs\rangle \simlt 5.5$, the optical-to-silicate extinction ratios of
these AGNs are substantially lower than that of the Galactic diffuse
interstellar medium (ISM) for which $A_V/\SilAbs\approx18.5$.  We argue that
the anomalously low $A_V/\SilAbs$ ratio could be due to the predominance of
larger grains in the AGN torus compared to that in the Galactic diffuse ISM. 
\end{abstract}


\section{Introduction}\label{sec:intro}
Active galaxies are eye-catching due to the ongoing energetic accretion of
nuclear material by their central supermassive black holes.  Observations have
shown that a large fraction of such activities are obscured by large columns of
dust and gas.  The unified model of active galactic nuclei \citep[AGNs;
see][]{Antoucci1993}, which features a torus-shaped structure of
obscuring material, successfully explains a large number of AGN observables.
Nevertheless, the properties of the dust in AGNs remain poorly understood
\citep[see][]{Li2007}.

An accurate knowledge of the dust extinction is crucial to recover the AGN
intrinsic spectra, and it also influences the estimation of important
parameters such as the AGN luminosity, black hole mass, and Eddington ratio. 
Numerous efforts have been made to expand such knowledge but have led to contradicting results: statistical
studies on the reddening of quasars suggest a steeply rising extinction curve
like (or even steeper than) that of the Small Magellanic Cloud (SMC), which
indicates the predominance of small grains \citep{Hall2002, Richard2003,
Hopkins2004, Glikman2012, Jiang2013},
while a flat or ``gray'' extinction curve (which varies little with wavelength)
has also been proposed, which suggests the richness of large grains
\citep{Gaskell2004, Czerny2004, Gaskell2007}.  We note that in deriving these
AGN extinction curves, one could suffer from the possible variations in the AGN
intrinsic spectra, orientation effects, or potentially biased sampling
\citep{Czerny2007}.

Alternatively, one could also probe the dust size and composition through the
dust infrared (IR) spectroscopy of AGNs.  As shown in \cite{Li2008},
\cite{Smith2010}, and \cite{Kohler2010}, the 9.7$\mum$ silicate emission
\citep{Hao2005c, Siebenmorgen2005, Sturm2005} and absorption \citep{Jaffe2004}
spectra of AGNs are diagnostic of the silicate composition and size. 

Dust size could also be probed through $A_V/\SilAbs$, the ratio of the visual
extinction $A_V$ to the 9.7$\mum$ silicate absorption depth $\SilAbs$
(\citealt{Gao2010}; Z. Shao et al.\ 2014, in preparation): for compact, spherical
silicate dust, $A_V/\SilAbs$ peaks at a grain size of $a\sim 0.2\mum$ and drops
precipitously with the increase of the dust size. 
In the local interstellar medium (ISM) of the Milky Way, $A_V/\SilAbs\approx
18.5\pm2$ (\citealt{Roche1984}; see Table~1 of \citealt{Draine-ARAA-2003} for a
summary).
In this Letter we report a considerably lower $A_V/\SilAbs$ ratio for 110 type 2
AGNs (Section \ref{sec:results}), with $A_V$ determined from the Balmer decrement
based on the Sloan Digital Sky Survey (SDSS) data, and $\SilAbs$ from the 9.7$\mum$ {\it
Spitzer}/InfraRed Spectroscope (IRS) absorption spectra (Section \ref{sec:sample}).
The anomalously low $A_V/\SilAbs$ ratio suggests the predominance of larger
grains in the AGN torus compared to the typical size of $a\sim0.1\mum$ in the
Galactic diffuse ISM (Section \ref{sec:discussion}). 
%


\section{Data and Measurements}\label{sec:sample}
\subsection{Sample}
We collect all type 2 AGNs from the \saga~sample (\spitzerirssdss Spectral
Atlas of Galaxies and AGN; L. Hao et al.\ 2014, in preparation). \saga~is a
{\it heterogeneous} collection of galaxies that have \spitzerirs
\citep{IRS2004} low-resolution spectra and SDSS spectroscopic observations
(Data Release 7; \citealt{Abazajian2009}) within a 3$^{\prime\prime}$ searching
radius.  The whole \saga~sample contains 139 type 1 AGNs, 114 type 2 AGNs, 241
star-forming (SF) galaxies, 103 AGN-SF composites, and 1 quiescent galaxy.
These classifications are made based on their SDSS optical spectra
\citep[see][]{Hao2005}.  Type 1 AGNs are those with broad \halpha\ emission
lines (with FWHM $>$\,1200$\km\s^{-1}$).  Type 2 AGNs are identified with the
typical ``Baldwin, Phillips \& Terlevich'' diagram \citep{BPT1981}. This
sample spans a redshift range of $z$\,$\simali$0.001--0.25, corresponding to a
physical size of $\simali$0.06--18\,kpc in the SDSS 3\arcsec~aperture.

The low-resolution mid-IR spectra of \spitzer are adopted from the Cornell
Atlas of {\it Spitzer}/IRS Sources \citep[CASSIS;][]{Lebouteiller2011}.  We use the
CASSIS v4 data. In this version, the intermediate products of the \spitzer
Science Center (SSC) pipeline release S18.7.0 are processed with the SMART
software with a dedicated spectral extraction pipeline developed by the CASSIS
team.  We further combine the spectra of the Short-Low (SL) and Long-Low (LL)
modules and scale the observed flux of LL to SL.  Spikes and edge fringes are
removed carefully.  The final order-combined mid-IR spectra span a wavelength
coverage of 5.2--38$\mum$ in the observer's frame, with a spectral resolution
of $\simali$60--127. We do not apply any aperture correction since the fiber
size of SDSS (3\arcsec) and slit width of \spitzerirs SL
($\simali$3$\arcsec$.6) are comparable. 

\subsection{Extinction Determination}
\label{sec:sample-measure}
The Balmer decrement can be used to trace the optical extinction: the
difference between the observed and intrinsic ratios of two hydrogen
recombination lines yields the amount of dust reddening. For the SDSS spectra,
we subtract the stellar continuum, decompose the \NII+\halpha~region and
\hbeta~region with multiple Gaussian components, and examine each fit by eye
with the aid of the $\chi^2$ statistic.  

We estimate the visual extinction from the measured ${\rm H\alpha/H\beta}$
ratios, assuming a ``screen'' dust configuration: 
\begin{equation}
\label{equ:av-screen2004}
A_V = 1.086\times \xi \times 
          \ln\left\{\frac{\rm (H\alpha/H\beta)_{\rm obs}}
          {\rm(H\alpha/H\beta)_{\rm int}}
          \right\} ~~,
\end{equation}
where
\begin{equation}
\label{equ:extcurv}
\xi \equiv \left(A_V/A_{\rm H\alpha}\right)
      \times \left(A_{\rm H\beta}/A_{\rm H\alpha} -1\right)^{-1} ~~.
\end{equation}
$A_{\rm H\alpha}$ and $A_{\rm H\beta}$ are the extinction at the H$\alpha$
(6562\,\AA) and H$\beta$ (4831\,\AA) bands, respectively. The parameter $\xi$ only
depends on how the extinction varies with wavelength: the grayer an extinction
curve is, the larger $\xi$ is. For a gray-type extinction curve such as
\cite{Gaskell2004}, $\xi\approx3.61$ (see Table \ref{tab:decrement}).
We adopt a standard ``Case B'' recombination for the intrinsic Balmer
decrement, i.e., ${\rm (H\alpha/H\beta)}_{\rm int}\approx 3.1$
\citep{Osterbrock2006}.

We define the silicate strength as
\begin{equation}
\label{equ:tau-screen}
S_{9.7} = -\ln\left(\frac{I_{\lambda\ast, {\rm obs}}}{I_{\lambda\ast, {\rm cont}}}\right),
\end{equation}
where $\lambda\ast$ is the wavelength of the 9.7$\mum$ silicate feature peak,
$I_{\lambda\ast, {\rm obs}}$ and $I_{\lambda\ast, {\rm cont}}$ are the
corresponding observed and continuum intensity, respectively.  We estimate the
silicate continuum using the interpolation method of \cite{Spoon2007} and the
PAHFIT decomposition method \citep{Smith-2007}. In general, the results based
on these two methods agree well. The final adopted 9.7$\mum$ silicate
strength is derived based on the interpolated continuum with the offset between
these two approaches as the error (See L. Hao et al. 2014, in preparation for details).

We {\it do not} apply Galactic extinction correction on any spectra, since none
of our objects locates along the Galactic plane. SDSS J031501.41+420208.9 is
the closest to the Galactic plane with its Galactic latitude at $b=-13^{\rm o}.3$.

\section{Results: Anomalously Low $A_V/\SilAbs$ 
           Ratios in Type 2 AGNs} \label{sec:results}
	  
\begin{figure}[htb]
\centering
\includegraphics[angle=0,width=1.0\hsize]{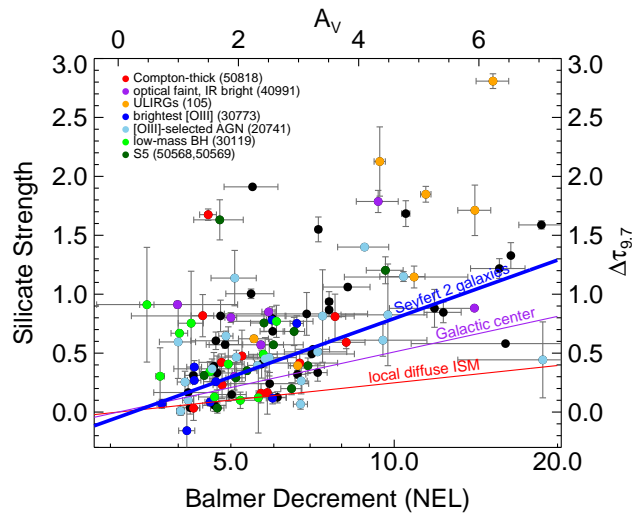}
\caption{\footnotesize 
    Observed Balmer decrements ${\rm (H\alpha/H\beta)}_{\rm obs}$ of the
    hydrogen narrow emission lines (NEL) vs. the 9.7$\mum$ silicate absorption
    strengths $S_{9.7}$ for our sample.  Also shown are the optical extinction
    $A_V$ (upper axis) derived from Equation \,(\ref{equ:av-screen2004}) and the
    9.7$\mum$ silicate absorption optical depth $\SilAbs$ (right axis) derived from
    Equation \,(\ref{equ:tau-bal}), assuming a ``screen'' dust geometry (i.e.,
    $\SilAbs=S_{9.7}$). Some of the data points obtained from different \spitzer
    programs are coded with different colors in order to show if the sources from
    any single program dominate or cluster in the figure (with program features and
    IDs shown in the legend).  The thick blue line is a linear fit of
    $A_V/\SilAbs\approx5.5\pm2.7$ (with an additional $y$-axis intercept
    $\simali$\,$-0.04\pm0.31$ and a reduced $\chi^2 \approx 43.2$). The red and
    purple solid lines plot $A_V/\SilAbs\approx 18.5$ for the local ISM of the
    Milky Way, and $A_V/\SilAbs\approx 9$ for the Galactic center, resectively.}
\label{fig:mir_ext5}
\end{figure}
%
For a ``screen'' dust configuration, $\SilAbs=S_{9.7}$.  Therefore, one would
expect a relation between the silicate absorption strength $S_{9.7}$ and
the Balmer decrement:
\begin{equation}
     S_{9.7} \approx 1.086\times\xi
              \times\left(A_V/\SilAbs\right)^{-1}
              \times \ln\left\{\frac{\rm (H\alpha/H\beta)_{\rm obs}}
              {\rm(H\alpha/H\beta)_{\rm int}}\right\} ~~.
\label{equ:tau-bal}
\end{equation}
Figure~\ref{fig:mir_ext5} shows the distribution of this sample on a plot of
the observed Balmer decrement $\left({\rm H\alpha/H\beta}\right)_{\rm obs}$
versus the observed silicate strength $S_{9.7}$.  A linear relation between
$S_{9.7}$ and $\ln\left({\rm H\alpha/H\beta}\right)_{\rm obs}$ is clearly seen
(albeit with a large scatter).  The slope of the correlation is
$1.086\times\xi\times\left(A_V/\SilAbs\right)^{-1}$. For an assumed dust
configuration, this slope is uniquely determined by the observed $S_{9.7}$ and
$\ln{\rm(H\alpha/H\beta)_{\rm obs}}$. Hence, for a given extinction law,
$A_V/\SilAbs\propto \xi$. 

Guided by Equation \,(\ref{equ:tau-bal}), we perform a linear 
weighted least-squares fit to the observationally determined parameters
$S_{9.7}$ and $\ln{\rm(H\alpha/H\beta)_{\rm obs}}$ shown in
Figure~\ref{fig:mir_ext5}.
The Spearman correlation test gives a coefficient  $r \approx 0.53$ with
significance $p\approx2.4\times 10^{-9}$, suggesting a moderate correlation
between $S_{9.7}$ and $\ln{\rm(H\alpha/H\beta)_{\rm obs}}$.
Assuming a Gaskell et al.-type extinction curve \citep{Gaskell2004} for which $\xi\approx3.61$,
we obtain $A_V/\SilAbs\approx5.5$. The bootstrap method is used to
estimate the errors of the fitted parameters with 5000 realizations (sampling
with replacement), and the resulting $1\sigma$ uncertainty for $A_V/\SilAbs$ is
2.4. In the fitting, we allow a small value of additional intercept to account
for the ${\rm(H\alpha/H\beta)_{\rm int}}$ term and the possible measurement
uncertainties.  The fitted additional intercept of $\sim$\,$-0.04\pm0.31$ is
indeed small.

\begin{figure}[htb]
\centering
\includegraphics[angle=0,width=1.0\hsize]{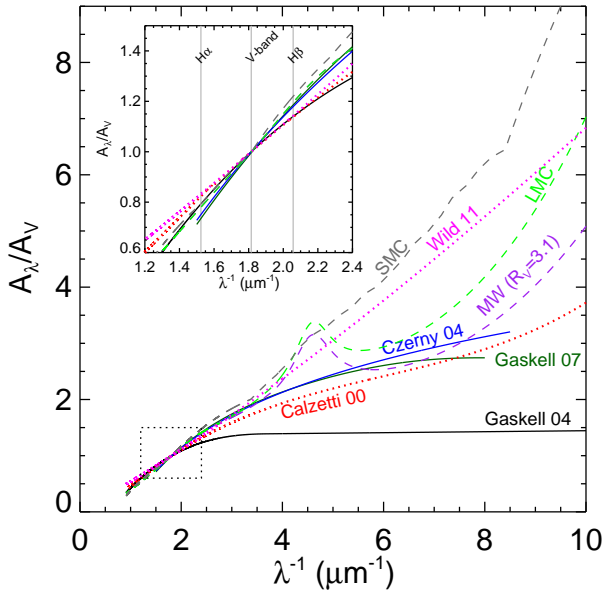}
\caption{\footnotesize 
	      Extinction curves commonly considered in AGN studies: flat or
	      gray curves of \cite{Gaskell2004}, \cite{Czerny2004}, and
	      \cite{Gaskell2007}. Also shown are the extinction curves of the
	      Milky Way diffuse ISM ($R_V=3.1$), the LMC, the SMC, and the steep
	      curves of \cite{Wild2011}. The attenuation curve of starburst
	      galaxies of \cite{Calzetti2000} is also plotted. The insert panel
	      highlights the extinction curves in the H$\alpha$, $V$, and
	      H$\beta$ bands.
              }
\label{fig:extcurv}
\end{figure}

The $A_V/\SilAbs\approx5.5 \pm 2.7$ ratio obtained above is significantly lower
than that of the Galactic diffuse ISM ($A_V/\SilAbs\approx 18.5\pm2$, see
review by \citealt{Draine-ARAA-2003}).  As shown in Figure~\ref{fig:extcurv}
and Table~\ref{tab:decrement}, the \cite{Gaskell2004} extinction curve is the
grayest and corresponds to the largest $\xi$ among all the extinction laws
commonly considered in AGN studies. As a result, any other extinction curve
would give an even lower $A_V/\SilAbs$.

The scatter of the correlation shown in Figure~\ref{fig:mir_ext5} is large (with a reduced
$\chi^2\approx43.2$). This is understandable since the intrinsic properties of
the dusty structures may be very different among type 2 AGNs. It
is also possible that there exists a systematic secondary effect in the
correlation causing the scatter (see discussion in Section \ref{sec:discussion}).
Nevertheless, it is clear, as can be seen from Figure \ref{fig:mir_ext5}, that most type 2 AGNs
lie above the lines given by the Galactic $A_V/\SilAbs$ ratios.

In our fitting, we exclude 4 out of the 114 type 2 AGNs in the \saga~sample:
3 of them have the Balmer decrement signal-to-noise ratio smaller than 3.0
and 1 source shows ${\rm (H\alpha/H\beta)}_{\rm obs} < 3.1$. Sample selection
is unlikely to bring significant bias on our result. Our 110 AGNs are selected
from 29 {\it Spitzer} programs whose scientific goals are substantially
different, without a single program dominating the sources or clustering in
Figure~\ref{fig:mir_ext5} (see the color-coded data points).

\begin{deluxetable}{lcccc}
\tablewidth{0pt}
\tablecaption{%
    \footnotesize
     \label{tab:decrement}
     $A_V/\SilAbs$ Derived from Various Extinction Curves
     }
\tablecolumns{5}
\tablehead{
    \colhead{Extinction Law} &
    \colhead{$A_{\rm H\beta}/A_{\rm H\alpha}$} &
    \colhead{$A_V/A_{\rm H\alpha}$} &
    \colhead{$\xi$} &
    \colhead{$A_V/\SilAbs$}
    }
\startdata
\cite{Gaskell2004}   &  1.34  & 1.21 & 3.61 & 5.5 \\
\cite{Gaskell2007}   &  1.61  & 1.37 & 2.25 & 3.4 \\
\cite{Czerny2004}    &  1.60  & 1.37 & 2.27 & 3.5 \\
\hline
MW $R_V$\,=\,3.1     &  1.51  & 1.29 & 2.44 & 3.7 \\
LMC Average          &  1.53  & 1.23 & 2.30 & 3.5 \\
SMC                  &  1.54  & 1.28 & 2.38 & 3.6 \\
\hline
\cite{Wild2011}      &  1.37  & 1.20 & 3.22 & 4.9 \\
\cite{Calzetti2000}  &  1.53  & 1.30 & 2.44 & 3.7
\enddata
\tablecomments{$A_V/\SilAbs \propto \xi$ where
      $\xi \equiv \left(A_V/A_{\rm H\alpha}\right)/
      \left(A_{\rm H\beta}/A_{\rm H\alpha} -1\right)$.
      }
\end{deluxetable}

\begin{figure}[htb]
\centering
\includegraphics[angle=0,width=0.7\hsize]{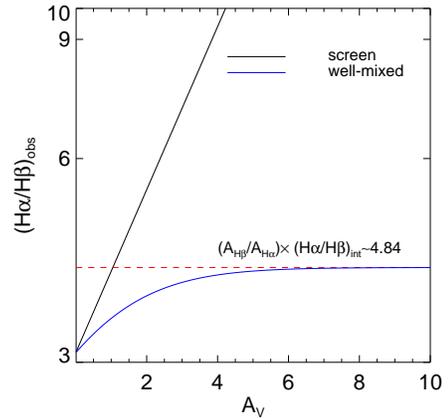}
\caption{\footnotesize 
	      Optical extinction $A_V$ derived from the observed Balmer
	      decrement ${\rm(H\alpha/H\beta)_{\rm obs}}$ with the assumption of
	      the gray extinction curve of \cite{Gaskell2004} in combination
	      with a ``screen'' (black solid line) or ``well-mixed'' (blue
	      solid line) dust configuration.  Also shown is the asymptotic
	      value of ${\rm(H\alpha/H\beta)_{\rm obs}} \approx \left(A_{\rm
	      H\beta}/A_{\rm H\alpha}\right) \times {\rm(H\alpha/H\beta)_{int}}
	      \approx 4.84$ for the ``screen'' geometry at
	      $A_V\rightarrow\infty$ (red dashed line; see
	      Equation\,(\ref{equ:av-mix})).  }
\label{fig:dust_geo}
\end{figure}

\section{Discussion}\label{sec:discussion}
The ``screen'' dust configuration is a simple assumption for the
obscuring AGN dust structure.  To examine the possible effects caused by the
deviation from a ``screen'' configuration, we consider two cases: (1) one
is an extreme case in which the dust is assumed to thoroughly mix with the
light source, but the dust distribution is still ``smooth,'' and (2) one is a
clumpy distribution of discrete dust clouds (e.g., see
\citealt{Nenkova2008a,Nenkova2008b,Nikutta2009}).

In a ``well-mixed'' geometry, the observed intensity $I_\lambda$ relates to the
intensity of the source $I_{\lambda}^{\rm o}$ through $I_\lambda =
I_{\lambda}^{\rm o} \{1-\exp\left(-\tau_\lambda\right)\}/\tau_\lambda$, where
$\tau_\lambda$ is the optical depth ($\tau_\lambda = A_\lambda/1.086$) at
wavelength $\lambda$ \citep{Mathis1972}. Therefore, the visual extinction can
be deduced from the observed Balmer decrement through
\begin{equation}
\label{equ:av-mix}
\frac{\left(\rm H\alpha/H\beta\right)_{\rm obs}}
        {\left(\rm H\alpha/H\beta\right)_{\rm int}}
= \left(\frac{A_{\rm H\beta}}{A_{\rm H\alpha}}\right)
    \times
    \frac{1-\exp\left\{-\left(A_V/1.086\right)
            \left(A_{\rm H\alpha}/A_V\right)\right\}}
            {1-\exp\left\{-\left(A_V/1.086\right)
            \left(A_{\rm H\beta}/A_V\right)\right\}} ~~.
\end{equation}
As shown in Figure~\ref{fig:dust_geo}, for a given extinction curve (which
specifies $A_{\rm H\beta}/A_{\rm H\alpha}$, $A_{\rm H\alpha}/A_V$, and $A_{\rm
H\beta}/A_V$), the ``mixed'' dust geometry always requires a larger optical
extinction $A_V$ to account for the same Balmer decrement $\left(\rm
H\alpha/H\beta\right)_{\rm obs}$. In this case, we will derive a higher
$A_V/\SilAbs$ compared to the ``screen'' scenario for the same Balmer decrement
and silicate data. Meanwhile, as can be seen in Equation~(\ref{equ:av-mix}) and in
Figure~\ref{fig:dust_geo}, the observed Balmer decrement cannot exceed
$\left(A_{\rm H\beta}/A_{\rm H\alpha}\right) \times\left(\rm
H\alpha/H\beta\right)_{\rm int}\approx 5.0$ (corresponding to $ A_{\rm
H\beta}/A_{\rm H\alpha}\approx 1.61$ for the extinction curve of
\citealt{Gaskell2007}), for any reasonable extinction curves considered in
Figure~\ref{fig:extcurv} and in Table~\ref{tab:decrement}.
However, as shown in Figure~\ref{fig:mir_ext5}, a large fraction of the sources
have $\left({\rm H\alpha/H\beta}\right)_{\rm obs}>5$ and they dominate the
overall trend of $A_V/\SilAbs$.  Therefore, the obscuring dust can not be in a
``well-mixed'' geometry.  For the 9.7$\mum$ silicate optical depth, one would
expect some degree of mixing since the silicate dust in the warm inner torus
will emit at 9.7$\mum$.  Taking into account the silicate emission, one should
derive a higher $\Delta\tau_{9.7}$ and hence an even lower $A_V/\SilAbs$. 

For a clumpy torus, we assume a Poisson distribution of discrete clouds.
Let $N$ be the average number of clouds along a radial equatorial line of
sight, and let $\tau_{\lambda,c}$ be the optical depth of a single cloud at
wavelength $\lambda$.  The observed intensity becomes $I_\lambda =
I_{\lambda}^{\rm o}
\exp\left\{-N\left[1-\exp\left(-\tau_{\lambda,c}\right)\right]\right\}$.
For the ``screen'' configuration, we can consider the ``screen'' as a smooth,
continuous distribution of $N$ clouds.  Hence the optical extinction and
silicate optical depth are $A_{V}=1.086\,N\tau_{V,c}$ and $\SilAbs
= N\Delta\tau_{9.7,c}$, respectively.  Consequently, the ratio of the clumpy
$\left(A_V/\SilAbs\right)_{\rm clum}$ to the screen
$\left(A_V/\SilAbs\right)_{\rm scrn}$ becomes $\left(A_V/\SilAbs\right)_{\rm
clum}/\left(A_V/\SilAbs\right)_{\rm scrn} =
\left\{1-\exp\left(-\tau_{V,c}\right)\right\}/\left\{1-\exp\left(-\tau_{9.7,c}\right)\right\}
\times \left\{\tau_{9.7,c}/\tau_{V,c}\right\}$.  It is easy to show
$\left(A_V/\SilAbs\right)_{\rm clum} < \left(A_V/\SilAbs\right)_{\rm scrn}$.
Therefore, the low optical-to-silicate extinction ratio of $A_V/\SilAbs\approx 5.5$
derived in Section~\ref{sec:results} indeed seems to already be in the high end.
Nevertheless, we note that in reality, the actual optical depth of
a clumpy geometry can be much more complicated than assumed above. Besides, a
clumpy geometry cannot produce silicate depths greater than $~0.5$ \citep[e.g.,
see][]{Levenson2007,Nikutta2009}, while a number of \saga~AGNs have silicate
absorption $S_{9.7}>0.5$. Thus, detailed modeling is needed to fully
address this issue.

\begin{figure}[htb]
\centering
\includegraphics[angle=0,width=0.7\hsize]{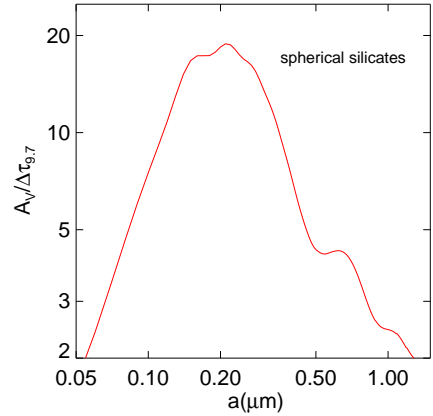}
\caption{\footnotesize 
       Optical to silicate extinction ratio ($A_V/\SilAbs$) of spherical silicate
       dust as a function of dust size.
}
\label{fig:grain_size}
\end{figure}

Dust scatters and absorbs light most effectively when its size $a$ is
comparable to the wavelength $\lambda$ of the light (i.e.,  $2\pi
a/\lambda\sim1$).  For the optical light, grains of $a\sim0.1\mum$ are an
effective scatter and absorber.  For larger grains (say, $a\simgt0.5\mum$),
they are less effective in extinguishing the optical light but become effective
in producing the 9.7$\mum$ silicate absorption feature (see Z. Shao et al.\
2014, in preparation).  The low $A_V/\SilAbs$ ratio of AGNs could readily be
explained in terms of dust larger than the submicro-sized interstellar dust. 
In Figure~\ref{fig:grain_size}, we show the optical-to-silicate extinction
ratios ($A_V/\SilAbs$) as a function of grain radii $a$ for compact, spherical
silicate grains calculated from Mie theory and the dielectric functions of
``astronomical silicates'' from \cite{Draine1984}.  It is seen that $A_V/\SilAbs$
peaks at $a\sim0.2\mum$ and rapidly decreases with the increase of $a$.  At
$a>0.44\mum$, we see $A_V/\SilAbs<5.5$.  Meanwhile, as shown in
Figure~\ref{fig:grain_size}, one could also achieve $A_V/\SilAbs<5.5$ with
$a<0.09\mum$. However, it is unlikely for these small grains to survive in the
hostile AGN circumnuclear environments, and the spectroscopic studies of
the 9.7$\mum$ silicate feature of AGNs all point to dust much larger than
$\sim0.1\mum$ \citep[see][]{Li2008, Kohler2010, Smith2010}. We note that
the observed high end of $A_V/\SilAbs\approx5.5$ does not necessarily imply
that the AGN dust grains are all larger than $a\approx0.44\mum$, since the
optical extinction is not exclusively contributed by silicate dust, although
the fractional contribution of carbon dust to $A_V$ is not known. 

There is evidence that suggests that the dust in the AGN torus is larger than
the dust in the ISM. The gray extinction derived for AGNs implies that the size
distribution of the dust in AGNs is skewed toward substantially large grains
(see Section~\ref{sec:intro}). Some infrared interferometry studies also show various
indications that the grain size around the torus might be larger \citep{Kishimoto2007,Kishimoto2009, Honig2013, Burtscher2013}.
\cite{Maiolino2001b} found that, for 16 of the 19 AGNs they studied, their
$E(B-V)/\NH$ and $A_V/\NH$ are significantly lower than the Galactic standard
values. \cite{Maiolino2001a} ascribed these reduced ratios in AGNs (often with
a solar or higher metallicity) to grain growth through coagulation in the dense
circumnuclear regions.  In this case, we could expect a dust size distribution
skewed toward large grains, leading a flat extinction curve \citep[but
see][, who argued that the X-ray absorption and optical extinction may occur in
distinct media]{Weingartner2002}. We note that the preferential destruction of
small dust by X-ray photons in the AGN torus could also lead to the
predominance of large dust and result in reduced $E(B-V)/\NH$ and $A_V/\NH$
ratios. 

A low $A_V/\SilAbs$ ratio could also be caused by a smaller abundance of
carbonaceous dust in AGNs than in the ISM. Toward the Galactic center, the 
optical-to-silicate extinction ratio is $A_V/\SilAbs\approx9$ \citep{Roche1985} which
is just $\simali$1/2 of that of the local ISM. \cite{Roche1985} attributed this
to the relative abundance of carbonaceous dust to silicate dust: in the
Galactic center there are more oxygen-rich stars which make silicate dust
\citep[but see][]{Gao2010}.
In AGN torus, if carbon dust is preferentially destroyed, one would also expect
a low $A_V/\SilAbs$ ratio since carbon dust only contributes to $A_V$ while
silicate dust contributes to both $A_V$ and $\SilAbs$. The harsh radiation in
AGNs is expected to destroy dust.  However, \cite{Perna2003} found that,
subject to X-ray/UV radiation, silicate grains would be preferentially
destroyed with respect to graphite.

\cite{Leipski2007} derived $A_V\approx 2.4$ (from the narrow line Balmer
decrement) and silicate optical depth $\SilAbs\approx0.34$ for a type 2 AGN
J12324114+1112587. This also implies a low optical-to-silicate extinction ratio
of $A_V/\SilAbs\approx 7.1$ compared with the Galactic value.
\cite{Leipski2007} explained this as that the optical extinction and the
silicate absorption may take place at different regions: the silicate
absorption mainly arises from the dust which is concentrated toward the center
of the galaxy, while the NLR emission comes from larger scales and is
additionally absorbed by ambient dust in the host galaxy. 
\cite{Goulding2012} explored the origin of the 9.7$\mum$ silicate
absorption in 20 nearby ($z < 0.05$) Compton-thick AGNs.  They accurately
measured the silicate absorption in archival low-resolution {\it Spitzer}/IRS
spectroscopy.  They found that $\simali$45\% of the sources have strong
silicate absorption with $S_{9.7} >0.5$.
Differing from \cite{Leipski2007}, \cite{Goulding2012} argued that the dominant
contribution to the observed silicate absorption is made by the dust located in
the host galaxy and not necessarily in a compact obscuring torus surrounding
the central engine. 
We note that the discussions on the low $A_V/\SilAbs$ ratio presented in this
work are based on the assumption that both the optical extinction and the
silicate absorption are from the torus. The optical extinction derived from the
NLR Balmer decrement may not be fully representative of the torus extinction
since the NLR may not be obscured by the entire torus. The
extinction can also come from the dust in hosts. However, if the hosts also
contribute to the mid-IR extinction, a low value of $A_V/\SilAbs$ ratio in AGNs
compared to the Galactic ISM can still be expected, since $\left[A_V\left({\rm
host}\right)+A_V\left({\rm torus}\right)\right]/\left[\SilAbs\left({\rm
host}\right)+\SilAbs\left({\rm torus}\right)\right] > A_V\left({\rm
torus}\right)/\SilAbs\left({\rm torus}\right)$. We call on further studies on
the origin of the optical and silicate extinction.
Nevertheless, if the optical extinction and the silicate absorption are indeed
from very different regions, one should not expect any
correlation between $A_V$ and $\SilAbs$. Mismatches in the
extinction regions traced by optical and silicate absorption may act as the
secondary effect, at most, in the correlation shown in
Figure~\ref{fig:mir_ext5}.

\section{Summary}\label{sec:summary}
We have determined $A_V/\SilAbs$, the ratio of the visual extinction to the
9.7$\mum$ silicate absorption optical depth, of 110 type 2 AGNs.  The mean
ratio of $A_V/\SilAbs\approx5.5$ is considerably lower than that of the local
ISM of the Milky Way.  We attribute this anomalously low $A_V/\SilAbs$ ratio of
AGNs to a dust size distribution skewed toward substantially large grains, as a
result of preferential destruction of small grains by X-ray/UV photons and/or
grain growth in the dense circumnuclear regions of AGNs.

\acknowledgments
We thank J.~Gao, B.W.~Jiang, Z.~Shao, Y.~Xie and the anonymous referee for
helpful suggestions. J.L. and L.H. are partially supported by the 973 Program
of China (2013CB834905, 2009CB824800), the Strategic Priority Research Program
``The Emergence of Cosmological Structures'' of Chinese Academy of Sciences
(XDB09030200), the Shanghai Pujiang Talents Program (10pj1411800) and NSFC
11073040. A.L. is supported in part by NSF AST-1311804 and NASA NNX14AF68G.
The Cornell Atlas of \spitzerirs Sources (CASSIS) is a product of the Infrared
Science Center at Cornell University, supported by NASA and JPL.



\begin{thebibliography}{}
\expandafter\ifx\csname natexlab\endcsname\relax\def\natexlab#1{#1}\fi

\bibitem[{{Abazajian} {et~al.}(2009){Abazajian}, {Adelman-McCarthy},
  {Ag{\"u}eros}, {Allam}, {Allende Prieto}, {An}, {Anderson}, {Anderson},
  {Annis}, {Bahcall}, \& et~al.}]{Abazajian2009}
{Abazajian}, K.~N., {Adelman-McCarthy}, J.~K., {Ag{\"u}eros}, M.~A., {et~al.}
  2009, \apjs, 182, 543

\bibitem[{{Antonucci}(1993)}]{Antoucci1993}
{Antonucci}, R. 1993, \araa, 31, 473

\bibitem[{{Baldwin} {et~al.}(1981){Baldwin}, {Phillips}, \&
  {Terlevich}}]{BPT1981}
{Baldwin}, J.~A., {Phillips}, M.~M., \& {Terlevich}, R. 1981, \pasp, 93, 5

\bibitem[{{Burtscher} {et~al.}(2013){Burtscher}, {Meisenheimer}, {Tristram},
  {Jaffe}, {H{\"o}nig}, {Davies}, {Kishimoto}, {Pott}, {R{\"o}ttgering},
  {Schartmann}, {Weigelt}, \& {Wolf}}]{Burtscher2013}
{Burtscher}, L., {Meisenheimer}, K., {Tristram}, K.~R.~W., {et~al.} 2013, \aap,
  558, A149

\bibitem[{{Calzetti} {et~al.}(2000){Calzetti}, {Armus}, {Bohlin}, {Kinney},
  {Koornneef}, \& {Storchi-Bergmann}}]{Calzetti2000}
{Calzetti}, D., {Armus}, L., {Bohlin}, R.~C., {et~al.} 2000, \apj, 533, 682

\bibitem[{{Czerny}(2007)}]{Czerny2007}
{Czerny}, B. 2007, in ASP Conf. Ser. 373, 
  The Central Engine of Active Galactic Nuclei, ed. L.~C. {Ho} \&
  J.-W. {Wang} (San Francisco, CA: ASP), 586

\bibitem[{{Czerny} {et~al.}(2004){Czerny}, {Li}, {Loska}, \&
  {Szczerba}}]{Czerny2004}
{Czerny}, B., {Li}, J., {Loska}, Z., \& {Szczerba}, R. 2004, \mnras, 348, L54

\bibitem[{{Draine}(2003)}]{Draine-ARAA-2003}
{Draine}, B.~T. 2003, \araa, 41, 241

\bibitem[{{Draine} \& {Lee}(1984)}]{Draine1984}
{Draine}, B.~T., \& {Lee}, H.~M. 1984, \apj, 285, 89

\bibitem[{{Gao} {et~al.}(2010){Gao}, {Jiang}, \& {Li}}]{Gao2010}
{Gao}, J., {Jiang}, B.~W., \& {Li}, A. 2010, EP\&S, 62, 63

\bibitem[{{Gaskell} \& {Benker}(2007)}]{Gaskell2007}
{Gaskell}, C.~M., \& {Benker}, A.~J. 2007, arXiv:0711.1013

\bibitem[{{Gaskell} {et~al.}(2004){Gaskell}, {Goosmann}, {Antonucci}, \&
  {Whysong}}]{Gaskell2004}
{Gaskell}, C.~M., {Goosmann}, R.~W., {Antonucci}, R.~R.~J., \& {Whysong}, D.~H.
  2004, \apj, 616, 147

\bibitem[{{Glikman} {et~al.}(2012){Glikman}, {Urrutia}, {Lacy}, {Djorgovski},
  {Mahabal}, {Myers}, {Ross}, {Petitjean}, {Ge}, {Schneider}, \&
  {York}}]{Glikman2012}
{Glikman}, E., {Urrutia}, T., {Lacy}, M., {et~al.} 2012, \apj, 757, 51

\bibitem[{{Goulding} {et~al.}(2012){Goulding}, {Alexander}, {Bauer}, {Forman},
  {Hickox}, {Jones}, {Mullaney}, \& {Trichas}}]{Goulding2012}
{Goulding}, A.~D., {Alexander}, D.~M., {Bauer}, F.~E., {et~al.} 2012, \apj,
  755, 5

\bibitem[{{Hall} {et~al.}(2002){Hall}, {Anderson}, {Strauss}, {York},
  {Richards}, {Fan}, {Knapp}, {Schneider}, {Vanden Berk}, {Geballe}, {Bauer},
  {Becker}, {Davis}, {Rix}, {Nichol}, {Bahcall}, {Brinkmann}, {Brunner},
  {Connolly}, {Csabai}, {Doi}, {Fukugita}, {Gunn}, {Haiman}, {Harvanek},
  {Heckman}, {Hennessy}, {Inada}, {Ivezi{\'c}}, {Johnston}, {Kleinman},
  {Krolik}, {Krzesinski}, {Kunszt}, {Lamb}, {Long}, {Lupton}, {Miknaitis},
  {Munn}, {Narayanan}, {Neilsen}, {Newman}, {Nitta}, {Okamura}, {Pentericci},
  {Pier}, {Schlegel}, {Snedden}, {Szalay}, {Thakar}, {Tsvetanov}, {White}, \&
  {Zheng}}]{Hall2002}
{Hall}, P.~B., {Anderson}, S.~F., {Strauss}, M.~A., {et~al.} 2002, \apjs, 141,
  267

\bibitem[{{Hao} {et~al.}(2005{\natexlab{a}}){Hao}, {Strauss}, {Tremonti},
  {Schlegel}, {Heckman}, {Kauffmann}, {Blanton}, {Fan}, {Gunn}, {Hall},
  {Ivezi{\'c}}, {Knapp}, {Krolik}, {Lupton}, {Richards}, {Schneider},
  {Strateva}, {Zakamska}, {Brinkmann}, {Brunner}, \& {Szokoly}}]{Hao2005}
{Hao}, L., {Strauss}, M.~A., {Tremonti}, C.~A., {et~al.} 2005{\natexlab{a}},
  \aj, 129, 1783

\bibitem[{{Hao} {et~al.}(2005{\natexlab{b}}){Hao}, {Spoon}, {Sloan},
  {Marshall}, {Armus}, {Tielens}, {Sargent}, {van Bemmel}, {Charmandaris},
  {Weedman}, \& {Houck}}]{Hao2005c}
{Hao}, L., {Spoon}, H.~W.~W., {Sloan}, G.~C., {et~al.} 2005{\natexlab{b}},
  \apjl, 625, L75

\bibitem[{{H{\"o}nig} {et~al.}(2013){H{\"o}nig}, {Kishimoto}, {Tristram},
  {Prieto}, {Gandhi}, {Asmus}, {Antonucci}, {Burtscher}, {Duschl}, \&
  {Weigelt}}]{Honig2013}
{H{\"o}nig}, S.~F., {Kishimoto}, M., {Tristram}, K.~R.~W., {et~al.} 2013, \apj,
  771, 87

\bibitem[{{Hopkins} {et~al.}(2004){Hopkins}, {Strauss}, {Hall}, {Richards},
  {Cooper}, {Schneider}, {Vanden Berk}, {Jester}, {Brinkmann}, \&
  {Szokoly}}]{Hopkins2004}
{Hopkins}, P.~F., {Strauss}, M.~A., {Hall}, P.~B., {et~al.} 2004, \aj, 128,
  1112

\bibitem[{{Houck} {et~al.}(2004){Houck}, {Roellig}, {van Cleve}, {Forrest},
  {Herter}, {Lawrence}, {Matthews}, {Reitsema}, {Soifer}, {Watson}, {Weedman},
  {Huisjen}, {Troeltzsch}, {Barry}, {Bernard-Salas}, {Blacken}, {Brandl},
  {Charmandaris}, {Devost}, {Gull}, {Hall}, {Henderson}, {Higdon}, {Pirger},
  {Schoenwald}, {Sloan}, {Uchida}, {Appleton}, {Armus}, {Burgdorf},
  {Fajardo-Acosta}, {Grillmair}, {Ingalls}, {Morris}, \& {Teplitz}}]{IRS2004}
{Houck}, J.~R., {Roellig}, T.~L., {van Cleve}, J., {et~al.} 2004, \apjs, 154,
  18

\bibitem[{{Jaffe} {et~al.}(2004){Jaffe}, {Meisenheimer}, {R{\"o}ttgering},
  {Leinert}, {Richichi}, {Chesneau}, {Fraix-Burnet}, {Glazenborg-Kluttig},
  {Granato}, {Graser}, {Heijligers}, {K{\"o}hler}, {Malbet}, {Miley},
  {Paresce}, {Pel}, {Perrin}, {Przygodda}, {Schoeller}, {Sol}, {Waters},
  {Weigelt}, {Woillez}, \& {de Zeeuw}}]{Jaffe2004}
{Jaffe}, W., {Meisenheimer}, K., {R{\"o}ttgering}, H.~J.~A., {et~al.} 2004,
  \nat, 429, 47

\bibitem[{{Jiang} {et~al.}(2013){Jiang}, {Zhou}, {Ji}, {Shu}, {Liu}, {Wang},
  {Dong}, {Bai}, {Wang}, \& {Wang}}]{Jiang2013}
{Jiang}, P., {Zhou}, H., {Ji}, T., {et~al.} 2013, \aj, 145, 157

\bibitem[{{Kishimoto} {et~al.}(2009){Kishimoto}, {H{\"o}nig}, {Antonucci},
  {Kotani}, {Barvainis}, {Tristram}, \& {Weigelt}}]{Kishimoto2009}
{Kishimoto}, M., {H{\"o}nig}, S.~F., {Antonucci}, R., {et~al.} 2009, \aap, 507,
  L57

\bibitem[{{Kishimoto} {et~al.}(2007){Kishimoto}, {H{\"o}nig}, {Beckert}, \&
  {Weigelt}}]{Kishimoto2007}
{Kishimoto}, M., {H{\"o}nig}, S.~F., {Beckert}, T., \& {Weigelt}, G. 2007,
  \aap, 476, 713

\bibitem[{{K{\"o}hler} \& {Li}(2010)}]{Kohler2010}
{K{\"o}hler}, M., \& {Li}, A. 2010, \mnras, 406, L6

\bibitem[{{Lebouteiller} {et~al.}(2011){Lebouteiller}, {Barry}, {Spoon},
  {Bernard-Salas}, {Sloan}, {Houck}, \& {Weedman}}]{Lebouteiller2011}
{Lebouteiller}, V., {Barry}, D.~J., {Spoon}, H.~W.~W., {et~al.} 2011, \apjs,
  196, 8

\bibitem[{{Leipski} {et~al.}(2007){Leipski}, {Haas}, {Meusinger},
  {Siebenmorgen}, {Chini}, {Drass}, {Albrecht}, {Wilkes}, {Huchra}, {Ott},
  {Cesarsky}, \& {Cutri}}]{Leipski2007}
{Leipski}, C., {Haas}, M., {Meusinger}, H., {et~al.} 2007, \aap, 464, 895

\bibitem[{{Levenson} {et~al.}(2007){Levenson}, {Sirocky}, {Hao}, {Spoon},
    {Marshall}, {Elitzur}, \& {Houck}}]{Levenson2007}
    {Levenson}, N.~A., {Sirocky}, M.~M., {Hao}, L., {et~al.} 2007, \apjl, 654, L45

\bibitem[{{Li}(2007)}]{Li2007}
{Li}, A. 2007, in ASP Conf. Ser. 373, The Central Engine of Active Galactic Nuclei, ed. L.~C. {Ho} \& J.-W.
  {Wang} (San Francisco, CA: ASP, 561

\bibitem[{{Li} {et~al.}(2008){Li}, {Shi}, \& {Li}}]{Li2008}
{Li}, M.~P., {Shi}, Q.~J., \& {Li}, A. 2008, \mnras, 391, L49


\bibitem[{{Maiolino} {et~al.}(2001{\natexlab{a}}){Maiolino}, {Marconi}, \&
  {Oliva}}]{Maiolino2001a}
{Maiolino}, R., {Marconi}, A., \& {Oliva}, E. 2001{\natexlab{a}}, \aap, 365, 37

\bibitem[{{Maiolino} {et~al.}(2001{\natexlab{b}}){Maiolino}, {Marconi},
  {Salvati}, {Risaliti}, {Severgnini}, {Oliva}, {La Franca}, \&
  {Vanzi}}]{Maiolino2001b}
{Maiolino}, R., {Marconi}, A., {Salvati}, M., {et~al.} 2001{\natexlab{b}},
  \aap, 365, 28

\bibitem[{{Mathis}(1972)}]{Mathis1972}
{Mathis}, J.~S. 1972, \apj, 176, 651

\bibitem[{{Nenkova} {et~al.}(2008{\natexlab{a}}){Nenkova}, {Sirocky},
  {Ivezi{\'c}}, \& {Elitzur}}]{Nenkova2008a}
{Nenkova}, M., {Sirocky}, M.~M., {Ivezi{\'c}}, {\v Z}., \& {Elitzur}, M.
  2008{\natexlab{a}}, \apj, 685, 147

\bibitem[{{Nenkova} {et~al.}(2008{\natexlab{b}}){Nenkova}, {Sirocky},
  {Nikutta}, {Ivezi{\'c}}, \& {Elitzur}}]{Nenkova2008b}
{Nenkova}, M., {Sirocky}, M.~M., {Nikutta}, R., {Ivezi{\'c}}, {\v Z}., \&
  {Elitzur}, M. 2008{\natexlab{b}}, \apj, 685, 160

\bibitem[{{Nikutta} {et~al.}(2009){Nikutta}, {Elitzur}, \&
  {Lacy}}]{Nikutta2009}
{Nikutta}, R., {Elitzur}, M., \& {Lacy}, M. 2009, \apj, 707, 1550

\bibitem[{{Osterbrock} \& {Ferland}(2006)}]{Osterbrock2006}
{Osterbrock}, D.~E., \& {Ferland}, G.~J. 2006, {Astrophysics of Gaseous Nebulae
  and Active Galactic Nuclei} (2nd ed.; Sausalito, CA: University Science Books)

\bibitem[{{Perna} {et~al.}(2003){Perna}, {Lazzati}, \& {Fiore}}]{Perna2003}
{Perna}, R., {Lazzati}, D., \& {Fiore}, F. 2003, \apj, 585, 775

\bibitem[{{Richards} {et~al.}(2003){Richards}, {Hall}, {Vanden Berk},
  {Strauss}, {Schneider}, {Weinstein}, {Reichard}, {York}, {Knapp}, {Fan},
  {Ivezi{\'c}}, {Brinkmann}, {Budav{\'a}ri}, {Csabai}, \&
  {Nichol}}]{Richard2003}
{Richards}, G.~T., {Hall}, P.~B., {Vanden Berk}, D.~E., {et~al.} 2003, \aj,
  126, 1131

\bibitem[{{Roche} \& {Aitken}(1984)}]{Roche1984}
{Roche}, P.~F., \& {Aitken}, D.~K. 1984, \mnras, 208, 481

\bibitem[{{Roche} \& {Aitken}(1985)}]{Roche1985}
---. 1985, \mnras, 215, 425

\bibitem[{{Siebenmorgen} {et~al.}(2005){Siebenmorgen}, {Haas}, {Kr{\"u}gel}, \&
  {Schulz}}]{Siebenmorgen2005}
{Siebenmorgen}, R., {Haas}, M., {Kr{\"u}gel}, E., \& {Schulz}, B. 2005, \aap,
  436, L5

\bibitem[{{Smith} {et~al.}(2010){Smith}, {Li}, {Li}, {K{\"o}hler}, {Ashby},
  {Fazio}, {Huang}, {Marengo}, {Wang}, {Willner}, {Zezas}, {Spinoglio}, \&
  {Wu}}]{Smith2010}
{Smith}, H.~A., {Li}, A., {Li}, M.~P., {et~al.} 2010, \apj, 716, 490

\bibitem[{{Smith} {et~al.}(2007){Smith}, {Draine}, {Dale}, {Moustakas},
  {Kennicutt}, {Helou}, {Armus}, {Roussel}, {Sheth}, {Bendo}, {Buckalew},
  {Calzetti}, {Engelbracht}, {Gordon}, {Hollenbach}, {Li}, {Malhotra},
  {Murphy}, \& {Walter}}]{Smith-2007}
{Smith}, J.~D.~T., {Draine}, B.~T., {Dale}, D.~A., {et~al.} 2007, \apj, 656,
  770

\bibitem[{{Spoon} {et~al.}(2007){Spoon}, {Marshall}, {Houck}, {Elitzur}, {Hao},
  {Armus}, {Brandl}, \& {Charmandaris}}]{Spoon2007}
{Spoon}, H.~W.~W., {Marshall}, J.~A., {Houck}, J.~R., {et~al.} 2007, \apjl,
  654, L49

\bibitem[{{Sturm} {et~al.}(2005){Sturm}, {Schweitzer}, {Lutz}, {Contursi},
  {Genzel}, {Lehnert}, {Tacconi}, {Veilleux}, {Rupke}, {Kim}, {Sternberg},
  {Maoz}, {Lord}, {Mazzarella}, \& {Sanders}}]{Sturm2005}
{Sturm}, E., {Schweitzer}, M., {Lutz}, D., {et~al.} 2005, \apjl, 629, L21

\bibitem[{{Weingartner} \& {Murray}(2002)}]{Weingartner2002}
{Weingartner}, J.~C., \& {Murray}, N. 2002, \apj, 580, 88

\bibitem[{{Wild} {et~al.}(2011){Wild}, {Groves}, {Heckman}, {Sonnentrucker},
  {Armus}, {Schiminovich}, {Johnson}, {Martins}, \& {Lamassa}}]{Wild2011}
{Wild}, V., {Groves}, B., {Heckman}, T., {et~al.} 2011, \mnras, 410, 1593

\end{thebibliography}
\end{document}